\documentclass{llncs}

\usepackage[T1]{fontenc}
\usepackage{amsmath,amssymb}
\usepackage{makeidx}
\usepackage{tabularx}
\usepackage{bussproofs}
\usepackage{graphicx}
\usepackage{url}

\def\atelierb{\textsf{Atelier~B}}
\def\bbook{\textsf{B}-Book}
\def\bcare{\textsf{BCARe}}
\def\bcoq{\textsf{BCoq}}
\def\bmth{\textsf{B}}

\def\cedric{\textsf{Cedric}}

\def\cnam{\textsf{Cnam}}
\def\coq{\textsf{Coq}}
\def\ensiie{\textsf{ENSIIE}}
\def\eventb{\textsf{Event-B}}
\def\inria{\textsf{Inria}}
\def\intelc{\textsf{Intel~Core~i5}}
\def\intelp{\textsf{Intel~Pentium~D~Xeon}}
\def\iproverm{\textsf{iProver~Modulo}}
\def\iprover{\textsf{iProver}}
\def\isahol{\textsf{Isabelle/HOL}}
\def\muscadet{\textsf{Muscadet}}
\def\ocaml{\textsf{OCaml}}
\def\siemens{\textsf{Siemens~IC-MOL}}

\def\szen{\textsf{Super~Zenon}}

\def\theorema{\textsf{Theorema}}
\def\verit{\textsf{veriT}}

\def\zenon{\textsf{Zenon}}

\def\bdots{\brsep ... \brsep}
\def\brsep{\;|\;}
\def\ex{\exists}
\def\fa{\forall}
\def\fctpartn{\shortmid\!\!\!\rightarrow}
\def\fctpart{~\fctpartn}
\def\funrule{fun}
\def\implies{\Rightarrow}
\def\overridet{~{-\mkern-20mu\lhd}\,}
\def\overriden{~{-\mkern-20mu\lhd}~}
\def\override{~\overriden~}
\def\pow{\mathbb{P}}
\def\predrule{pred}
\def\substleftn{{-\mkern-14mu\lhd}}
\def\substleft{~\substleftn~}
\def\substrightn{{-\mkern-14mu\rhd}}
\def\substright{~\substrightn~}
 
\EnableBpAbbreviations 
\newcommand{\UICm}[1]{\UIC{$#1$}}
\newcommand{\AXCm}[1]{\AXC{$#1$}}
\newcommand{\BICm}[1]{\BIC{$#1$}}

\newcommand{\RLm}[1]{\RL{$#1$}}

\newenvironment{ack}{\bigskip{}\noindent{}\emph{\ackname}}{}

\begin{document}

\title{Tableaux Modulo Theories using Superdeduction}
\titlerunning{Tableaux Modulo Theories using Superdeduction}

\author{Mélanie~Jacquel\inst{1} \and Karim~Berkani\inst{1} \and
David~Delahaye\inst{2} \and Catherine~Dubois\inst{3}}
\authorrunning{Mélanie~Jacquel et al.}
\tocauthor{Mélanie~Jacquel, David~Delahaye, Catherine~Dubois, and Karim~Berkani}

\institute{\siemens{}, Châtillon, France,\\
\email{Melanie.Jacquel@siemens.com}\\
\email{Karim.Berkani@siemens.com} \and
\cedric{}/\cnam{}/\inria{}, Paris, France,\\
\email{David.Delahaye@cnam.fr} \and
\cedric{}/\ensiie{}/\inria{}, Évry, France,\\
\email{dubois@ensiie.fr}}

\maketitle

\begin{abstract}
We propose a method that allows us to develop tableaux modulo theories using the
principles of superdeduction, among which the theory is used to enrich the
deduction system with new deduction rules. This method is presented in the
framework of the \zenon{} automated theorem prover, and is applied to the set
theory of the \bmth{} method. This allows us to provide another prover to
\atelierb{}, which can be used to verify \bmth{} proof rules in particular. We
also propose some benchmarks, in which this prover is able to automatically
verify a part of the rules coming from the database maintained by
\siemens{}. Finally, we describe another extension of \zenon{} with
superdeduction, which is able to deal with any first order theory, and provide a
benchmark coming from the TPTP library, which contains a large set of first
order problems.

\keywords{Automated Deduction, Tableaux, Superdeduction, \zenon{}, Set Theory,
\bmth{} Method, First Order Theories.}
\end{abstract}


\section{Introduction}

Reasoning modulo a theory like arithmetic in first order logic may appear as a
complex task, since even simple formulas become difficult to be proved (using
natural deduction or sequent calculus), and the corresponding proofs are hardly
readable. This is actually due to the fact that Gentzen's deductive systems,
even if more elaborate than systems à la Hilbert, consist of low-level languages
for deduction. Over the last few years, several approaches have been developed
to palliate this problem, and for instance, deduction modulo~\cite{DA03} and
superdeduction~\cite{BA07}, which respectively focus on the computational and
deductive parts of a theory, can be considered as steps toward high-level
deductive languages.

Superdeduction has been adapted for usual deductive systems, such as natural
deduction or sequent calculus (see~\cite{BA07}, for example). In these systems,
we have to deal with some common permutability problems of automated proof
search, since superdeduction systems are in fact embedding a part of compiled
automated deduction. As a consequence, superdeduction can be naturally
integrated into automated deduction methods, and we propose to do so in the
framework of the tableau method (this corresponds to an alternative approach to
the work described in~\cite{RB04}, where a tableau method for deduction modulo
is proposed).

The integration of superdeduction into the tableau method is actually motivated
by an experiment that is managed by \siemens{} regarding the verification of
\bmth{} proof rules~\cite{JA13}. The \bmth{} method~\cite{B-Book}, or \bmth{}
for short, allows engineers to develop software with high guarantees of
confidence; more precisely it allows them to build correct by design
software. \bmth{} is a formal method based on set theory and theorem proving,
and which relies on a refinement-based development process. The \atelierb{}
environment~\cite{Atelier-B} is a platform that supports \bmth{} and offers,
among other tools, both automated and interactive provers. In practice, to
ensure the global correctness of formalized applications, the user must
discharge proof obligations. These proof obligations may be proved
automatically, but otherwise, they have to be handled manually either by using
the interactive prover, or by adding new proof rules that the automated prover
can exploit. These new proof rules can be seen as axioms and must be verified by
other means, otherwise the global correctness may be endangered.

In~\cite{JA13}, we develop an approach based on the use of the \zenon{}
automated theorem prover~\cite{Zenon}, which relies on classical first order
logic with equality and applies the tableau method as proof search. In this
context, the choice of \zenon{} is strongly influenced by its ability of
producing comprehensible proof traces under the form of \coq{} proofs~\cite{Coq}
in particular. The method used in this approach consists in first normalizing
the formulas to be proved, in order to obtain first order logic formulas
containing only the membership set operator, and then calling \zenon{} on these
new formulas. This experiment gives satisfactory results in the sense that it
can prove a significant part of the rules coming from the database maintained by
\siemens{} (we can deal with about 1,400~rules, 1,100~of which can be proved
automatically, over a total of 5,300~rules). However, this approach is not
complete (after the normalization, \zenon{} proves the formulas without any
axiom of set theory, while some instantiations may require to be normalized),
and suffers from efficiency issues (due to the preliminary normalization). To
deal with these problems, the idea developed in this paper is to integrate the
axioms and constructs of the \bmth{} set theory into the \zenon{} proof search
method by means of superdeduction rules. This integration can be concretely
achieved thanks to the extension mechanism offered by \zenon{}, which allows us
to extend its core of deductive rules to match specific requirements.

In this paper, we also propose another extension of \zenon{} with
superdeduction, which is able to deal with any first order theory, and which
must be seen as a generalization of the extension of \zenon{} with
superdeduction and for the \bmth{} set theory that has been described
previously. This other extension of \zenon{} has been developed as a tool called
\szen{}~\cite{Super-Zenon}, where each theory is analyzed to determine the
axioms which can be turned into superdeduction rules, and these superdeduction
rules are automatically computed on the fly to enrich the deductive kernel of
\zenon{}.

The paper is organized as follows: in Sec.~\ref{sec:rwork}, we start by briefly
reviewing some related work; in Secs.~\ref{sec:ratio} and~\ref{sec:sded},
we then respectively introduce superdeduction and present the computation of
superdeduction rules from axioms in the framework of the tableau method used by
\zenon{}; next, we explain, in Sec.~\ref{sec:bset}, the superdeduction rules
corresponding to the set theory which the \bmth{} method relies on, and
describe, in Sec.~\ref{sec:bench}, the implementation of our extension of
\zenon{} for the \bmth{} set theory and provide some comparative benchmarks
concerning the verification of \bmth{} proof rules coming from the database
maintained by \siemens{}; thereafter, in Sec.~\ref{sec:first}, we present the
other extension of \zenon{} with superdeduction, which is able to deal with any
first order theory, and also provide a comparative benchmark coming from the
TPTP library, which contains a large set of first order problems in particular;
finally, in Sec.~\ref{sec:sauto}, we discuss more generally the use of
superdeduction for proof search in axiomatic theories.


\section{Related Work}
\label{sec:rwork}

In the framework of verification of \bmth{} (or \eventb{}) proof obligations,
and in addition to \bmth{} dedicated provers, such as the main prover of
\atelierb{}, some similar work has been done to use external provers, relying on
either SMT (Satisfiability Modulo Theories) solvers like \verit{}~\cite{DD13},
or proof assistants like \coq{}~\cite{CA05} or \isahol{}~\cite{SA11}. However,
\verit{} does not provide proof traces, and automation is poor in the considered
proof assistants, even though dedicated tactics are offered. More generally,
there are also some provers covering set theory, such as \muscadet{}~\cite{DP02}
or \theorema{}~\cite{WW06}, but they do not provide traces either.

Regarding the ability of reasoning modulo theories, there exist some alternative
approaches, such as the technology of SMT solvers. This technology is based on a
SAT solver and a combination of several decision procedures for several
theories. There are several methods to combine the considered decision
procedures, such as the Nelson-Oppen~\cite{NO79} or Shostak~\cite{RS84}
methods. The approach used by SMT solvers is actually restrictive in the sense
that the considered theories have to be decidable. To preserve this
decidability, the combination methods also impose some restrictions over the
theories to be combined, which must have disjoint signatures. The techniques of
deduction modulo~\cite{DA03} and superdeduction~\cite{BA07}, proposed in this
paper in particular, are more general approaches to integrate theories into
proof search methods since there is no constraint over the theories to be
integrated, which may be decidable or not.


\section{Rationale for Superdeduction}
\label{sec:ratio}

Deduction modulo~\cite{DA03} focuses on the computational part of a theory,
where axioms are transformed into rewrite rules, which induces a congruence over
propositions, and where reasoning is performed modulo this
congruence. Superdeduction~\cite{BA07} is actually a variant of deduction
modulo, where axioms are used to enrich the deduction system with new deduction
rules, which are called superdeduction rules. For example, considering the
definition of inclusion in set theory
$\fa{}a\fa{}b\;((a\subseteq{}b)\Leftrightarrow{}
(\fa{}x\;(x\in{}a\implies{}x\in{}b)))$, the proof of $A\subseteq{}A$ in sequent
calculus has the following form:

\begin{center}
\AXCm{}\RLm{\mathrm{Ax}}
\UICm{\ldots{},x\in{}A\vdash{}A\subseteq{}A,x\in{}A}\RLm{{\implies}\mathrm{R}}
\UICm{\ldots{}\vdash{}A\subseteq{}A,x\in{}A\implies{}x\in{}A}
\RLm{\fa{}\mathrm{R}}
\UICm{\ldots{}\vdash{}A\subseteq{}A,\fa{}x\;(x\in{}A\implies{}x\in{}A)}
\AXCm{}\RLm{\mathrm{Ax}}
\UICm{\ldots{},A\subseteq{}A\vdash{}A\subseteq{}A}\RLm{{\implies}\mathrm{L}}
\BICm{\ldots{},
(\fa{}x\;(x\in{}A\implies{}x\in{}A))\implies{}A\subseteq{}A\vdash{}
A\subseteq{}A}\RLm{\land{}\mathrm{L}}
\UICm{A\subseteq{}A\Leftrightarrow{}
(\fa{}x\;(x\in{}A\implies{}x\in{}A))\vdash{}A\subseteq{}A}
\RLm{\fa{}\mathrm{L}\times{}2}
\UICm{\fa{}a\fa{}b\;((a\subseteq{}b)\Leftrightarrow{}
(\fa{}x\;(x\in{}a\implies{}x\in{}b)))\vdash{}A\subseteq{}A}\DP
\end{center}

In deduction modulo, the axiom of inclusion can be seen as a computation rule
and therefore replaced by the rewrite rule
$a\subseteq{}b\rightarrow{}\fa{}x\;(x\in{}a\implies{}x\in{}b)$. The previous
proof is then transformed as follows:

\begin{center}
\AXCm{}\RLm{\mathrm{Ax}}
\UICm{x\in{}A\vdash{}x\in{}A}\RLm{{\implies}\mathrm{R}}
\UICm{\vdash{}x\in{}A\implies{}x\in{}A}
\RLm{\fa{}\mathrm{R}\mbox{, }
A\subseteq{}A\rightarrow{}\fa{}x\;(x\in{}A\implies{}x\in{}A)}
\UICm{\vdash{}A\subseteq{}A}\DP
\end{center}

It can be noticed that the obtained proof is much simpler than the one completed
using sequent calculus. In addition to simplicity, deduction modulo also allows
us for unbounded proof size speed-up~\cite{GB11a}.

Superdeduction proposes to go further than deduction modulo precisely when the
considered axiom defines a predicate $P$ with an equivalence
$\forall{}\bar{x}\;(P\Leftrightarrow{}\varphi{})$. While deduction modulo
replaces the axiom by a rewrite rule, superdeduction adds to this transformation
the decomposition of the connectives occurring in this definition. This
corresponds to an extension of Prawitz's folding (resp. unfolding)
rules~\cite{DP65}, where a maximum of connectives of the definition are
introduced (resp. eliminated). Superdeduction may be seen as the alliance of
deduction modulo with focusing, which is a technique initially introduced in the
framework of linear logic~\cite{JMA92}. For the axiom of inclusion, the proposed
superdeduction rule is therefore the following (there is also a corresponding
left rule):

\begin{center}
\AXCm{\Gamma{},x\in{}a\vdash{}x\in{}b,\Delta{}}
\RLm{\mathrm{IncR}\mbox{, }x\not\in{}\Gamma{},\Delta{}}
\UICm{\Gamma{}\vdash{}a\subseteq{}b,\Delta{}}\DP
\end{center}

Hence, proving $A\subseteq{}A$ with this new rule can be performed as follows:

\begin{center}
\AXCm{}\RLm{\mathrm{Ax}}
\UICm{x\in{}A\vdash{}x\in{}A}\RLm{\mathrm{IncR}}
\UICm{\vdash{}A\subseteq{}A}\DP
\end{center}

This new proof is not only simpler and shorter than in deduction modulo, but
also follows a natural human reasoning scheme usually used in mathematics
(in the same vein, see~\cite{JA13a} for more details about how superdeduction
allows us to recover intuition from automated proofs).


\section{From Axioms to Superdeduction Rules}
\label{sec:sded}

As mentioned previously, reasoning modulo a theory in a tableau method using
superdeduction requires to generate new deduction rules from some axioms of the
theory. The axioms which can be considered for superdeduction are of the form
$\forall{}\bar{x}\;(P\Leftrightarrow{}\varphi{})$, where $P$ is atomic. This
specific form of axiom allows us to introduce an orientation of the axiom from
$P$ to $\varphi{}$, and we introduce the notion of proposition rewrite rule
(this notion appears in~\cite{BA07}, from which we borrow the following
definition and notation):

\begin{definition}[Proposition Rewrite Rule]
The notation $R:P\rightarrow{}\varphi{}$ denotes the axiom
$\forall{}\bar{x}\;(P\Leftrightarrow{}\varphi{})$, where $R$ is the name of the
rule, $P$ an atomic proposition, $\varphi{}$ a proposition, and $\bar{x}$ the
free variables of $P$ and $\varphi{}$.
\end{definition}

It should be noted that $P$ may contain first order terms and therefore that
such an axiom is not just a definition. For instance,
$x\in{}a\cap{}b\rightarrow{}x\in{}a\land{}x\in{}b$ (where $a\cap{}b$ is a first
order term) is a proposition rewrite rule.

As said in the introduction, one of our main objectives is to develop a proof
search procedure for the set theory of the \bmth{} method using the \zenon{}
automated theorem prover~\cite{Zenon}. In the following, we will thus consider
the tableau method used by \zenon{} as the framework in which superdeduction
rules will be generated from proposition rewrite rules.

The proof search rules of \zenon{} are described in detail in~\cite{Zenon} and
summarized in Fig.~\ref{fig:zenon} (for the sake of simplification, we have
omitted the unfolding and extension rules), where $\epsilon$ is Hilbert's
operator ($\epsilon(x).P(x)$ means some $x$ that satisfies $P(x)$, and is
considered as a term), capital letters are used for metavariables, and $R_r$,
$R_s$, $R_t$, and $R_{ts}$ are respectively reflexive, symmetric, transitive,
and transitive-symmetric relations (the corresponding rules also apply to the
equality in particular). As hinted by the use of Hilbert's operator, the
$\delta{}$-rules are handled by means of $\epsilon{}$-terms rather than using
Skolemization. What we call here metavariables are often named free variables in
the tableau-related literature; they are not used as variables as they are never
substituted. The proof search rules are applied with the normal tableau method:
starting from the negation of the goal, apply the rules in a top-down fashion to
build a tree. When all branches are closed (i.e. end with an application of a
closure rule), the tree is closed, and this closed tree is a proof of the
goal. Note that this algorithm is applied in strict depth-first order: we close
the current branch before starting work on another branch. Moreover, we work in
a non-destructive way: working on one branch will never change the formulas of
any other branch. We divide these rules into five distinct classes to be used
for a more efficient proof search. This extends the usual sets of rules dealing
with $\alpha,\beta,\delta,\gamma$-formulas and closure ($\odot$) with the
specific rules of \zenon{}. We list below the five sets of rules and their
elements:

$$\begin{array}{|c|l|}
\hline 
~ \alpha ~ & \alpha_{\neg\vee},\alpha_\land, \alpha_{\neg\implies}, 
\alpha_{\neg\neg},\mathbf{\neg_{\text{refl}}}\\
\hline
\beta &
\beta_\vee,\beta_{\neg\land},\beta_\implies,\beta_{\Leftrightarrow},
\beta_{\neg\Leftrightarrow}, \text{\predrule, \funrule, sym, trans}\ast\\
\hline
\delta & \delta_\ex,\delta_{\neg\fa}\\
\hline
\gamma &
\gamma_{\fa M},\gamma_{\neg\ex M},\gamma_{\fa\text{inst}},
\gamma_{\neg\ex\text{inst}}\\
\hline
\odot & \odot_\top,\odot_\bot,\odot,\odot_{r},\odot_{s}\\
\hline 
\end{array}$$

\medskip{}

where ``trans$\ast$'' gathers all the transitivity rules.

\begin{figure}[htbp]
\framebox[\textwidth][c]
{\parbox{\textwidth}
{\small
\hspace{0.2cm}\underline{Closure and Cut Rules}
\begin{center}
\begin{tabular}{c@{\hspace{1cm}}c@{\hspace{1cm}}c}
\AXCm{\bot}\RL{$\odot_\bot$}\UICm{\odot}\DP &
\AXCm{\neg \top}\RL{$\odot_{\neg\top}$}\UICm{\odot}\DP &
\AXCm{~}\RL{cut}\UICm{P\brsep \neg P}\DP\\\\
\AXCm{\neg R_r(t,t)}\RL{$\odot_r$}\UICm{\odot}\DP &
\AXCm{P}\AXCm{\neg P}\RL{$\odot$} \BICm{\odot}\DP &
\AXCm{R_s(a,b)}\AXCm{\neg R_s(b,a)}
\RL{$\odot_s$}\BICm{\odot}\DP
\end{tabular}
\end{center}
\hspace{0.2cm}\underline{Analytic Rules}
\begin{center}
\begin{tabular}{c@{\hspace{1cm}}c@{\hspace{1cm}}c}
\AXCm{\neg\neg P}\RL{$\alpha_{\neg\neg}$}
\UICm{P}\DP &
\AXCm{P\Leftrightarrow Q}\RL{$\beta_{\Leftrightarrow}$}
\UICm{\neg P, \neg Q\brsep P,Q}\DP &
\AXCm{\neg (P\Leftrightarrow Q)}\RL{$\beta_{\neg\Leftrightarrow}$}
\UICm{\neg P, Q\brsep P,\neg Q}\DP\\\\
\AXCm{P\land Q}\RL{$\alpha_{\land}$}\UICm{P,Q}\DP &
\AXCm{\neg (P\vee Q)}\RL{$\alpha_{\neg \vee}$}\UICm{\neg P, \neg Q}\DP &
\AXCm{\neg (P\implies Q)}\RL{$\alpha_{\neg\implies}$}\UICm{P,\neg Q}\DP\\\\
\AXCm{P\vee Q}\RL{$\beta_{\vee}$}\UICm{P\brsep Q}\DP &
\AXCm{\neg (P\land Q)}
\RL{$\beta_{\neg\land}$}\UICm{\neg P\brsep \neg Q}\DP &
\AXCm{P\implies Q}\RL{$\beta_{\implies}$}\UICm{\neg P\brsep Q}\DP\\\\
\multicolumn{3}{c}{
\begin{tabular}{cc}
\AXCm{\ex x\; P(x)}\RL{$\delta_\ex$}
\UICm{P(\epsilon(x).P(x))}\DP
~~~&~~~ 
\AXCm{\neg\fa x\; P(x)}\RL{$\delta_{\neg\fa}$}
\UICm{\neg P(\epsilon(x).\neg P(x))}\DP
\end{tabular}}
\end{tabular}
\end{center}
\hspace{0.2cm}\underline{$\gamma$-Rules}
\begin{center}
\begin{tabular}{c@{\hspace{1cm}}c}
\AXCm{\fa x\; P(x)} \RL{$\gamma_{\fa M}$}
\UICm{P(X)} \DP &
\AXCm{\neg \ex x\; P(x)} \RL{$\gamma_{\neg\ex M}$}
\UICm{\neg P(X)} \DP\\\\
 \AXCm{\fa x\; P(x)} \RL{$\gamma_{\fa \text{inst}}$}
\UICm{P(t)} \DP &
\AXCm{\neg \ex x\; P(x)} \RL{$\gamma_{\neg\ex \text{inst}}$}
\UICm{\neg P(t)} \DP\\
\end{tabular}
\end{center}
\hspace{0.2cm}\underline{Relational Rules}
\begin{center}
\begin{tabular}{cc}
\AXCm{P(t_1,...,t_n)}
\AXCm{\neg P(s_1,..,s_n)} \RL{\predrule}
\BICm{t_1\neq s_1\bdots t_n\neq s_n}\DP
~~~&~~~
\AXCm{f(t_1,...,t_n)\neq f(s_1,...,s_n)}
\RL{\funrule}
\UICm{t_1\neq s_1\bdots t_n\neq s_n}\DP\\\\
\AXCm{R_s(s,t)} \AXCm{\neg R_s(u,v) }
\RL{sym}
\BICm{t\neq u\brsep s\neq v}\DP &
\AXCm{\neg R_r(s,t)}
\RL{$\neg_{\text{refl}}$}
\UICm{s\neq t}\DP\\\\
\multicolumn{2}{c}{
\AXCm{R_t(s,t)} \AXCm{\neg R_t(u,v)}
\RL{trans}
\BICm{u\neq s, \neg R_t(u,s)\brsep t\neq v, \neg R_t(t,v)}\DP}\\\\
\multicolumn{2}{c}{
\AXCm{R_{ts}(s,t)} \AXCm{\neg R_{ts}(u,v)}
\RL{transsym}
\BICm{v\neq s, \neg R_{ts}(v,s)\brsep t\neq u, \neg R_{ts}(t,u)}\DP}\\\\
\multicolumn{2}{c}{
\AXCm{s = t} \AXCm{\neg R_t(u,v)}
\RL{transeq}
\BICm{u\neq s,\neg R_t(u,s)\brsep \neg R_t(u,s),\neg R_t(t,v)\brsep t\neq
v, \neg R_t(t,v)}\DP}\\\\
\multicolumn{2}{c}{
\AXCm{s = t} \AXCm{\neg R_{ts}(u,v)}
\RL{transeqsym}
\BICm{v\neq s,\neg R_{ts}(v,s)\brsep \neg R_{ts}(v,s),\neg R_{ts}(t,u)
\brsep t\neq u, \neg R_{ts}(t,u)}\DP}\\
\end{tabular}
\end{center}}}
\caption{Proof Search Rules of \zenon{}}
\label{fig:zenon}
\end{figure}

Let us now describe how the computation of superdeduction rules for \zenon{} is
performed from a given proposition rewrite rule.

\begin{definition}[Computation of Superdeduction Rules]
\label{def:sded-gen}
Let ${\cal S}$ be a set of rules composed by the subset of the proof search
rules of \zenon{} formed of the closure rules, the analytic rules, as well as
the $\gamma_{\fa M}$ and $\gamma_{\neg\ex M}$ rules. Given a proposition rewrite
rule $R:P\rightarrow{}\varphi{}$, two superdeduction rules (a positive one $R$
and a negative one $\neg{}R$) are generated in the following way:

\begin{enumerate}
\item To get the positive rule $R$, initialize the procedure with the formula
$\varphi{}$. Next, apply the rules of ${\cal S}$ until there is no open leaf
anymore on which they can be applied. Then, collect the premises and the
conclusion, and replace $\varphi$ by $P$ to obtain the positive rule $R$.

\item To get the negative rule $\neg{}R$, initialize the procedure with the
formula $\neg{}\varphi{}$. Next, apply the rules of ${\cal S}$ until there is no
open leaf anymore on which they can be applied. Then, collect the premises and
the conclusion, and replace $\neg{}\varphi$ by $\neg{}P$ to obtain the negative
rule $\neg{}R$.
\end{enumerate}

If the rule $R$ (resp. $\neg{}R$) involves metavariables, an instantiation rule
$R_\mathrm{inst}$ (resp. $\neg{}R_\mathrm{inst}$) is added, where one or several
metavariables can be instantiated.
\end{definition}

Integrating these new deduction rules to the proof search rules of \zenon{} is
trivially correct as they are generated from a subset of the rules of \zenon{},
while the considered subset of rules only consists of reversible rules, which is
a necessary condition for completeness.

Let us illustrate the computation of superdeduction rules from a proposition
rewrite rule with the example of the set inclusion.

\begin{example}[Set Inclusion]
From the definition of the set inclusion, we introduce the proposition rewrite
rule
$\mathrm{Inc}:a\subseteq{}b\rightarrow{}\fa{}x\;(x\in{}a\implies{}x\in{}b)$, and
the corresponding superdeduction rules $\mathrm{Inc}$ and $\neg{}\mathrm{Inc}$
are generated as follows:

\begin{center}
\begin{tabular}{cp{0.5cm}c}
\AXCm{\fa{}x\;(x\in{}a\implies{}x\in{}b)}\RLm{\gamma_{\fa M}}
\UICm{X\in{}a\implies{}X\in{}b}\RLm{\beta_\implies}
\UICm{X\not\in{}a\brsep X\in{}b}\DP &&

\AXCm{\neg\fa{}x\;(x\in{}a\implies{}x\in{}b)}\RLm{\delta_{\neg\fa}}
\UICm{\neg(\epsilon_x\in{}a\implies{}\epsilon_x\in{}b)}
\RLm{\alpha_{\neg\implies}}
\UICm{\epsilon_x\in{}a,\epsilon_x\not\in{}b}\DP
\end{tabular}
\end{center}

where $\epsilon_x=\epsilon(x).\neg(x\in{}a\implies{}x\in{}b)$.

The resulting superdeduction rules are then the following:

\begin{center}
\begin{tabular}{cp{0.5cm}cp{0.5cm}c}

\AXCm{a\subseteq{}b}\RLm{\mathrm{Inc}}
\UICm{X\not\in{}a\brsep X\in{}b}\DP &&

\AXCm{a\subseteq{}b}\RLm{\mathrm{Inc}_\mathrm{inst}}
\UICm{t\not\in{}a\brsep t\in{}b}\DP &&

\AXCm{a\not\subseteq{}b}\RLm{\neg{}\mathrm{Inc}}
\UICm{\epsilon_x\in{}a,\epsilon_x\not\in{}b}\DP
\end{tabular}
\end{center}
\end{example}


\section{Superdeduction Rules for the \bmth{} Set Theory}
\label{sec:bset}

The purpose of this section is to build a superdeduction system for the \bmth{}
set theory, which can be used for the verification of proof rules of
\atelierb{}, such as the rules coming from the database maintained by \siemens{}
(see Sec.~\ref{sec:bench}). To do so, the idea is to apply the computational
algorithm of superdeduction rules described in Sec.~\ref{sec:sded} to the
several axioms of the \bmth{} set theory.

\subsection{The \bmth{} Set Theory}

As said in the introduction, the \bmth{} method~\cite{B-Book} aims to assist
experts to develop certified software. The initial step is defined with abstract
properties of a model. Several steps of property refinement are then applied
until the release of the complete software. A refinement step is characterized
by adding details on the software behavior under construction. For each step,
generated proof obligations must be demonstrated to ensure the global
correctness of the formalization.

The \bmth{} method is based on a typed set theory. There are two rule systems:
one for demonstrating that a formula is well-typed, and one for demonstrating
that a formula is a logical consequence of a set of axioms. The main aim of the
type system is to avoid inconsistent formulas, such as Russell's paradox for
example. The \bmth{} proof system is based on a sequent calculus with equality.
Six axioms define the basic operators and the extensionality which, in turn,
defines the equality of two sets. In addition, the other operators ($\cup{}$,
$\cap{}$, etc.) are defined using the previous basic
ones. Fig.~\ref{fig:bset-def} gathers a part of the axioms and constructs of the
\bmth{} set theory, where $\mathrm{BIG}$ is an infinite set (mostly only used to
build natural numbers in the foundational theory). In this figure, we only
consider the four first axioms of the \bmth{} set theory, as we do not need the
two remaining axioms in the rules that we want to verify (see
Sec.~\ref{sec:bench}). Regarding functions and due to space restrictions, we
only present the notion of partial function, even though we can deal with the
other function constructs in our superdeduction system. Compared
to~\cite{B-Book}, all type information has been removed from the axioms and
constructs. This can be done since typechecking is performed before proving a
formula, and the pieces of type information dispersed throughout the axioms and
constructs are useless when building a proof (see~\cite{AM02} for details
regarding the modularity between the type and proof systems). Finally, we do not
deal with substitution explicitly (see the axiom for comprehension sets, for
example), as we do not intend to manage rules involving explicit substitutions
(see Sec.~\ref{sec:bench}).

\begin{figure}[tb]
\framebox[12.2cm][c]
{\parbox{12.2cm}
{\small
\hspace{0.2cm}\underline{Axioms}
\begin{flushleft}
$\begin{array}{p{1cm}l}
& (x,y)\in{}a\times{}b\Leftrightarrow{}x\in{}a\land{}y\in{}b\\
& a\in{}\pow{}(b)\Leftrightarrow{}\fa{}x\;(x\in{}a\implies{}x\in{}b)\\
& x\in{}\{~y~|~P(y)~\}\Leftrightarrow{}P(x)\\
& a=b\Leftrightarrow{}\fa{}x\;(x\in{}a\Leftrightarrow{}x\in{}b)
\end{array}$
\end{flushleft}
\hspace{0.2cm}\underline{Derived Constructs}
\begin{center}
$\begin{array}{l@{\hspace{1cm}}l}
a\subseteq{}b\Leftrightarrow{}a\in{}\pow{}(b) &
a\subset{}b\Leftrightarrow{}a\subseteq{}b\land{}a\neq{}b\\
a\cup{}b\triangleq{}\{~x~|~x\in{}a\vee{}x\in{}b~\} &
a\cap{}b\triangleq{}\{~x~|~x\in{}a\land{}x\in{}b~\}\\
a-b\triangleq{}\{~x~|~x\in{}a\land{}x\not\in{}b~\} &
\emptyset{}\triangleq{}\mathrm{BIG}-\mathrm{BIG}\\
\multicolumn{2}{l}{\{~e_1,\ldots,e_n~\}\triangleq{}
\{~x~|~x=e_1~\}\cup{}\ldots\cup{}\{~x~|~x=e_n~\}}
\end{array}$
\end{center}
\hspace{0.2cm}\underline{Binary Relation Constructs: First Series}
\begin{center}
$\begin{array}{l@{\hspace{1cm}}l}
a\leftrightarrow{}b\triangleq{}\pow{}(a\times{}b) &
a^{-1}\triangleq{}\{~(y,x)~|~(x,y)\in{}a~\}\\
\mathrm{dom}(a)\triangleq{}\{~x~|~\ex{}y\;(x,y)\in{}a~\} &
\mathrm{ran}(a)\triangleq{}\mathrm{dom}(a^{-1})\\
a;b\triangleq{}\{~(x,z)~|~\ex{}y\;((x,y)\in{}a\land{}(y,z)\in{}b~\} &
a\circ{}b\triangleq{}b;a\\
\mathrm{id}(a)\triangleq{}\{~(x,y)~|~(x,y)\in{}a\times{}a\land{}x=y~\}\\
a\lhd{}b\triangleq{}\mathrm{id}(a);b &
a\rhd{}b\triangleq{}a;\mathrm{id}(b)\\
a\substleft{}b\triangleq{}(\mathrm{dom}(b)-a)\lhd{}b &
a\substright{}b\triangleq{}a\rhd{}(\mathrm{ran}(a)-b)
\end{array}$
\end{center}
\hspace{0.2cm}\underline{Binary Relation Constructs: Second Series}
\begin{center}
$\begin{array}{l@{\hspace{1cm}}l}
a[b]\triangleq{}\mathrm{ran}(b\lhd{}a) &
a\override{}b\triangleq{}(\mathrm{dom}(b)\substleft{}a)\cup{}b\\
a\otimes{}b\triangleq{}\{~(x,(y,z))~|~(x,y)\in{}a\land{}(x,z)\in{}b~\}\\
\mathrm{prj}_1(a,b)\triangleq{}(\mathrm{id}(a)\otimes{}(a\times{}b))^{-1}
& \mathrm{prj}_2(a,b)\triangleq{}((b\times{}a)\otimes{}\mathrm{id}(b))^{-1}\\
a~||~b\triangleq{}\{~(x,y),(z,t)~|~(x,z)\in{}a\land{}(y,t)\in{}b~\}
\end{array}$
\end{center}
\hspace{0.2cm}\underline{Function Constructs: First Series}
\begin{flushleft}
$\begin{array}{p{1cm}l}
& a\fctpart{}b\triangleq{}\{~t~|~t\in{}a\leftrightarrow{}b\land{}
\fa{}(x,y,z)\;((x,y)\in{}t\land{}(x,z)\in{}t\implies{}y=z)~\}
\end{array}$
\end{flushleft}}}
\caption{Axioms and Constructs of the \bmth{} Set Theory}
\label{fig:bset-def}
\end{figure}

\subsection{Generating the Superdeduction Rules}

To generate the superdeduction rules corresponding to the axioms and constructs
defined in Fig.~\ref{fig:bset-def}, we use the algorithm described in
Def.~\ref{def:sded-gen} of Sec.~\ref{sec:sded}, and we must therefore identify
the several proposition rewrite rules. On the one hand, the axioms are all of
the form $P_i\Leftrightarrow{}Q_i$, and the associated proposition rewrite rules
are $R_i:P_i\rightarrow{}Q_i$, where each axiom is oriented from left to
right. On the other hand, the constructs are expressed by the definitions
$E_i\triangleq{}F_i$, where $E_i$ and $F_i$ are expressions, and the
corresponding proposition rewrite rules are
$R_i:x\in{}E_i\rightarrow{}x\in{}F_i$; if $E_i$ and $F_i$ represent relations,
the proposition rewrite rules are $R_i:(x,y)\in{}E_i\rightarrow{}(x,y)\in{}F_i$,
where we consider that relations are defined by means of ordered pairs. The
superdeduction rules generated from this set of proposition rewrite rules are
described in Figs.~\ref{fig:bset-sded1} and~\ref{fig:bset-sded2}. Due to space
restrictions, we do not include the definitions of $\subseteq$ (as well as
$\subset$), and $\circ$ in these figures, but their rules can be respectively
deduced from the rules of $\pow$ and ``$;$''. For the same reasons, we do not
describe the instantiation rules associated with each rule involving
metavariables. It should also be noted that the computation of these
superdeduction rules goes further than the one proposed in Sec.~\ref{sec:sded},
since given a proposition rewrite rule $R:P\rightarrow{}Q$, we apply to $Q$ not
only all the rules considered by Definition~\ref{def:sded-gen}, but also the new
generated superdeduction rules (except the rules for the extensional equality,
in order to benefit from the dedicated rules of \zenon{} for equality) whenever
applicable.

\begin{figure}[htbp]
\framebox[12.2cm][c]
{\parbox{12.2cm}
{\small
\hspace{0.2cm}\underline{Rules for Axioms}
\begin{center}
\begin{tabular}{c@{\hspace{0.5cm}}c@{\hspace{0.5cm}}c}
\AXCm{(x,y){}\in{}a\times{}b}\RLm{\times}
\UICm{x{}\in{}a,y{}\in{}b}\DP &
\AXCm{a\in{}\pow{}(b)}\RLm{\pow}
\UICm{X\not\in{}a\brsep{}X\in{}b}\DP &
\AXCm{x\in{}\{~y~|~P(y)~\}}\RLm{\{|\}}
\UICm{P(x)}\DP\\\\
\AXCm{(x,y)\not\in{}a\times{}b}\RLm{\lnot{}\times}
\UICm{x{}\not\in{}a\brsep{}y\not\in{}b}\DP &
\begin{tabular}{c}
\AXCm{a\not\in{}\pow{}(b)}\RLm{\lnot{}\pow}
\UICm{\epsilon_x\in{}a,\epsilon_x\not\in{}b}\DP\smallskip{}\\
{\scriptsize with $\epsilon_x=\epsilon(x).\lnot{}(x\in{}a\implies{}x\in{}b)$}
\end{tabular}&
\AXCm{x\not\in{}\{~y~|~P(y)~\}}\RLm{\lnot{}\{|\}}
\UICm{\lnot{}P(x)}\DP\\\\
\multicolumn{3}{c}{
\begin{tabular}{c@{\hspace{0.5cm}}c}
\AXCm{a=b}\RLm{=}
\UICm{X\not\in{}a,X\not\in{}b\brsep{}X\in{}a,X\in{}b}\DP &
\begin{tabular}{c}
\AXCm{a\neq{}b}\RLm{\neq}
\UICm{\epsilon_x\not\in{}a,\epsilon_x\in{}b\brsep{}\epsilon_x\in{}a,
\epsilon_x\not\in{}b}\DP\smallskip{}\\
{\scriptsize with
$\epsilon_x=\epsilon(x).\lnot{}(x\in{}a\Leftrightarrow{}x\in{}b)$}
\end{tabular}\\
\end{tabular}}
\end{tabular}
\end{center}
\hspace{0.2cm}\underline{Rules for Derived Constructs}
\begin{center}
\begin{tabular}{c@{\hspace{0.5cm}}c@{\hspace{0.5cm}}c}
\AXCm{x\in{}a\cup{}b}\RLm{\cup}
\UICm{x\in{}a\brsep{}x\in{}b}\DP &
\AXCm{x\in{}a\cap{}b}\RLm{\cap}
\UICm{x\in{}a,x\in{}b}\DP &
\AXCm{x\in{}a-b}\RLm{-}
\UICm{x\in{}a,x\not\in{}b}\DP\\\\
\AXCm{x\not\in{}a\cup{}b}\RLm{\lnot{}\cup}
\UICm{x\not\in{}a,x\not\in{}b}\DP &
\AXCm{x\not\in{}a\cap{}b}\RLm{\lnot{}\cap}
\UICm{x\not\in{}a\brsep{}x\not\in{}b}\DP &
\AXCm{x\not\in{}a-b}\RLm{\lnot{}-}
\UICm{x\not\in{}a\brsep{}x{}\in{}b}\DP\\\\
\AXCm{x\in{}\{~e_1,\ldots,e_n~\}}\RLm{\{\}}
\UICm{x=e_1\brsep{}\ldots{}\brsep{}x=e_1}\DP &
\AXCm{x\not\in{}\{~e_1,\ldots,e_n~\}}\RLm{\lnot{}\{\}}
\UICm{x\neq{}e_1,\ldots,x\neq{}e_n}\DP &
\AXCm{x\in{}\emptyset{}}\RLm{\emptyset}
\UICm{\odot{}}\DP\\\\
\end{tabular}
\end{center}
\hspace{0.2cm}\underline{Rules for Binary Relation Constructs: First Series}
\begin{center}
\begin{tabular}{c@{\hspace{0.5cm}}c@{\hspace{0.5cm}}c}
\AXC{$(x,y){}\in{}a^{-1}$}\RL{$a^{-1}$}
\UIC{$(y,x){}\in{}a$}\DP &
\begin{tabular}{c}
\AXC{$x{}\in{}\mathrm{dom}(a)$}\RL{$\mathrm{dom}$}
\UIC{$(x,\epsilon_y){}\in{}a$}\DP\smallskip{}\\
{\scriptsize with $\epsilon_y=\epsilon(y).((x,y){}\in{}a)$}
\end{tabular} &
\begin{tabular}{c}
\AXC{$y{}\in{}\mathrm{ran}(a)$}\RL{$\mathrm{ran}$}
\UIC{$(\epsilon_x,y){}\in{}a$}\DP\smallskip{}\\
{\scriptsize with $\epsilon_x=\epsilon(x).((x,y){}\in{}a)$}
\end{tabular}\\\\
\AXC{$(x,y){}\not\in{}a^{-1}$}\RL{$\lnot{}a^{-1}$}
\UIC{$(y,x){}\not\in{}a$}\DP &
\AXC{$x{}\not\in{}\mathrm{dom}(a)$}\RL{$\lnot{}\mathrm{dom}$}
\UIC{$(x,Y){}\not\in{}a$}\DP &
\AXC{$y{}\not\in{}\mathrm{ran}(a)$}\RL{$\lnot{}\mathrm{ran}$}
\UIC{$(X,y){}\not\in{}a$}\DP\\
\end{tabular}
\end{center}
\begin{center}
\begin{tabular}{c@{\hspace{0.5cm}}c@{\hspace{0.5cm}}c}
\multicolumn{3}{c}{
\begin{tabular}{c@{\hspace{1cm}}c}
\begin{tabular}{c}
\AXC{$(x,z)\in{}a;b$}\RL{$;$}
\UIC{$(x,\epsilon_y)\in{}a$, $(\epsilon_y,z)\in{}b$}\DP\smallskip{}\\
{\scriptsize with $\epsilon_y=\epsilon(y).((x,y){}\in{}a\land{}(y,z){}\in{}b)$}
\end{tabular}&
\AXC{$(x,z)\not\in{}a;b$}\RL{$\lnot{};$}
\UIC{$(x,Y)\not\in{}a\brsep{}(Y,z)\not\in{}b$}\DP\\\\
\AXC{$(x,y)\in{}\mathrm{id}(a)$}\RL{$id$}
\UIC{$x=y$, $x\in{}a$, $y\in{}a$}\DP &
\AXC{$(x,y)\not\in{}\mathrm{id}(a)$}\RL{$\lnot{}id$}
\UIC{$x\neq{}y\brsep{}x\not\in{}a\brsep{}y\not\in{}a$}\DP\\\\
\AXC{$(x,y)\in{}a\lhd{}b$}\RL{$\lhd$}
\UIC{$(x,y)\in{}b$, $x\in{}a$}\DP &
\AXC{$(x,y)\not\in{}a\lhd{}b$}\RL{$\lnot{}\lhd$}
\UIC{$(x,y)\not\in{}b\brsep{}x\not\in{}a$}\DP
\end{tabular}}
\end{tabular}
\end{center}}}
\caption{Superdeduction Rules for the \bmth{} Set Theory (Part~1)}
\label{fig:bset-sded1}
\end{figure}

\begin{figure}[htbp]
\framebox[12.2cm][c]
{\parbox{12.2cm}
{\small
\begin{center}
\begin{tabular}{c@{\hspace{0.5cm}}c@{\hspace{0.5cm}}c}
\AXC{$(x,y)\in{}a\substleft{}b$}\RL{$\substleftn$}
\UIC{$(x,y)\in{}b$, $x\not\in{}a$}\DP&
\AXC{$(x,y)\in{}a\rhd{}b$}\RL{$\rhd$}
\UIC{$(x,y)\in{}a$, $y\in{}b$}\DP&
\AXC{$(x,y)\in{}a\substright{}b$}\RL{$\substrightn$}
\UIC{$(x,y)\in{}a$, $y\not\in{}b$}\DP\\\\
\AXC{$(x,y)\not\in{}a\substleft{}b$}\RL{$\lnot{}\substleftn$}
\UIC{$(x,y)\not\in{}b\brsep{}x\in{}a$}\DP&
\AXC{$(x,y)\not\in{}a\rhd{}b$}\RL{$\lnot{}\rhd$}
\UIC {$(x,y)\not\in{}a\brsep{}y\not\in{}b$}\DP&
\AXC{$(x,y)\not\in{}a\substright{}b$}\RL{$\lnot{}\substrightn$}
\UIC {$(x,y)\not\in{}a\brsep{}y\in{}b$}\DP\\\\
\end{tabular}
\end{center}
\hspace{0.2cm}\underline{Rules for Binary Relation Constructs: Second Series}
\begin{center}
\begin{tabular}{c@{\hspace{0.1cm}}c}
\begin{tabular}{c}
\AXC{$y\in{}a[b]$}\RL{$a[b]$}
\UIC{$\epsilon_x\in{}b,(\epsilon_x,y)\in{}a$}\DP\smallskip{}\\
{\scriptsize with $\epsilon_x=\epsilon(x).(x\in{}b\land{}(x,y){}\in{}a)$}
\end{tabular} &
\AXC{$(x,y){}\in{}a\override{}b$}\RL{$\overriden$}
\UIC{$(x,y)\in{}a$, $(x,Y)\not\in{}b\brsep{}(x,y)\in{}b$}\DP\\\\
\AXC{$y\not\in{}a[b]$}\RL{$\lnot{}a[b]$}
\UIC{$X\not\in{}b\brsep{}(X,y)\not\in{}a$}\DP &
\begin{tabular}{c} 
\AXC{$(x,y){}\not\in{}a\override{}b$}\RL{$\lnot{}\overriden$}
\UIC{$(x,y)\not\in{}a$, $(x,y)\not\in{}b\brsep{}(x,\epsilon_y)\in{}b$,
$(x,y)\not\in{}b$}\DP\smallskip{}\\
{\scriptsize with $\epsilon_y=\epsilon(y).((x,y){}\in{}b)$}
\end{tabular}
\end{tabular}
\end{center}
\begin{center}
\begin{tabular}{c@{\hspace{0.1cm}}c}
\AXC{$(x,(y,z))\in{}a\otimes{}b$}\RL{$\otimes$}
\UIC{$(x,y)\in{}a$, $(x,z)\in{}b$}\DP &
\AXC{$((x,y),z)\in{}\mathrm{prj}_1(a,b)$}\RL{$\mathrm{prj}_1$}
\UIC{$x\in{}a$, $y\in{}b$, $z\in{}a$, $z=x$}\DP\\\\
\AXC{$(x,(y,z))\in{}a\otimes{}b$}\RL{$\lnot{}\otimes$}
\UIC {$(x,y)\not\in{}a\brsep{}(x,z)\not\in{}b$}\DP &
\AXC{$((x,y),z)\not\in{}\mathrm{prj}_1(a,b)$}\RL{$\lnot{}\mathrm{prj}_1$}
\UIC{$x\not\in{}a\brsep{}y\not\in{}b\brsep{}z\not\in{}a\brsep{}z\neq{}x$}\DP\\\\
\AXC{$((x,y),z)\in{}\mathrm{prj}_2(a,b)$}\RL{$\mathrm{prj}_2$}
\UIC{$x\in{}a$, $y\in{}b$, $z\in{}b$, $z=y$}\DP &
\AXC{$(x,y),(z,t))\in{}a~||~b$}\RL{$||$}
\UIC{$(x,z)\in{}a$, $(y,t)\in{}b$}\DP\\\\
\AXC{$((x,y),z)\not\in{}\mathrm{prj}_2(a,b)$}\RL{$\lnot{}\mathrm{prj}_2$}
\UIC{$x\not\in{}a\brsep{}y\not\in{}b\brsep{}z\not\in{}b\brsep{}z\neq{}y$}\DP &
\AXC{$(x,y),(z,t))\not\in{}a~||~b$}\RL{$\lnot{}||$}
\UIC{$(x,z)\not\in{}a\brsep{}(y,t)\not\in{}b$}\DP 
\end{tabular}
\end{center}
\hspace{0.2cm}\underline{Rules for Functions}
\begin{center}
\begin{tabular}{c}
\AXC{$a\in{}b\fctpart{}c$}\RL{$\fctpart$}
\UIC{$\begin{array}{c}
T\notin{}a,(X,Y)\not\in{}a\brsep{}
T\notin{}a,(X,Z)\not\in{}a\brsep{}
T\notin{}a,Y=Z\brsep{}\\
\Delta,(X,Y)\notin{}a\brsep{}
\Delta,(X,Z)\notin{}a\brsep{} 
\Delta,Y=Z
\end{array}$}\DP\\
{\scriptsize with 
$\Delta\equiv{}T=(\epsilon_x,\epsilon_y),\epsilon_x\in{}b,\epsilon_y\in{}c$ }\\ 
{\scriptsize where 
$\epsilon_x=\epsilon(x).\ex{}y\;(T=(x,y)\land{}(x,y)\in{}b\times{}c)$}\\
{\scriptsize 
$\epsilon_y=
\epsilon(y).(T=(\epsilon_x,y)\land{}(\epsilon_x,y)\in{}b\times{}c)$ }\\\\
\AXC{$a\not\in{}b\fctpart{}c$}\RL{$\lnot{}\fctpartn$}
\UIC{$\begin{array}{c}
\epsilon_x'\in{}a,\epsilon_x'\neq{}(Y,Z)\brsep{}
\epsilon_x'\in{}a,Y\notin{}b\brsep{}
\epsilon_x'\in{}a,Z\notin{}c\brsep{}\\
(\epsilon_x,\epsilon_y)\in{}a$, 
$(\epsilon_x,\epsilon_z)\in{}a$, 
$\epsilon_y\neq{}\epsilon_z
\end{array}$}\DP\\
{\scriptsize with 
$\epsilon_x'=\epsilon(x).\lnot{}(x\in{}a\Rightarrow{}x\in{}b\times{}c)$}\\
{\scriptsize
$\epsilon_x=\epsilon(x).\lnot{}(\fa{}y\fa{}z\;
((x,y)\in{}a\land{}(x,z)\in{}a\Rightarrow{}y=z))$}\\
{\scriptsize
$\epsilon_y=\epsilon(y).\lnot{}(\fa{}z\;
((\epsilon_x,y)\in{}a\land{}(\epsilon_x,z)\in{}a\Rightarrow{}y=z))$}\\
{\scriptsize
$\epsilon_z=\epsilon(z).\lnot{}
((\epsilon_x,\epsilon_y)\in{}a\land{}
(\epsilon_x,z)\in{}a\Rightarrow{}\epsilon_y=z)$}\\
\end{tabular}
\end{center}}}
\caption{Superdeduction Rules for the \bmth{} Set Theory (Part~2)}
\label{fig:bset-sded2}
\end{figure}

Regarding the superdeduction rules for the constructs, we can notice that given
a definition $E\triangleq{}F$, the generation of superdeduction rules is
performed from the proposition rewrite rule $R:x\in{}E\rightarrow{}x\in{}F$, and
we can therefore wonder if completeness can still be ensured. In particular, we
have to show that $P(E)\Leftrightarrow{}P(F)$, where $P$ is a predicate
symbol. This is actually possible using the extensional equality (rules $=$ and
$\neq$) together with the substitution rules (rules $\mathrm{pred}$ and
$\mathrm{fun}$) as follows:

$$\rootAtTop
\AXCm{\begin{array}{c}\vdots{}\\
\odot{}\end{array}}
\UICm{\epsilon_x\not\in{}E,\epsilon_x\in{}F}
\AXCm{\begin{array}{c}\vdots{}\\
\odot{}\end{array}}
\UICm{\epsilon_x\in{}E,\epsilon_x\not\in{}F}\RLm{\neq{}}
\BICm{E\neq{}F}\RLm{\mathrm{pred}}
\UICm{\neg{}P(E),P(F)}
\AXCm{\begin{array}{c}\vdots{}\\
\odot{}\end{array}}\UICm{P(E),\neg{}P(F)}\RLm{\beta_{\neg{}\Leftrightarrow}}
\BICm{\neg{}(P(E)\Leftrightarrow{}P(F))}\DP$$

where $\epsilon_x=\epsilon(x).\lnot{}(x\in{}E\Leftrightarrow{}x\in{}F)$, and
where the superdeduction rules for $R:x\in{}E\rightarrow{}x\in{}F$ can be
respectively applied on $\epsilon_x\in{}E$ and $\epsilon_x\not\in{}E$.

\subsection{Dealing with Relations}
\label{subsec:rel}

As mentioned previously, the superdeduction rules for relations are generated
from proposition rewrite rules of the form
$R:(x,y)\in{}E\rightarrow{}(x,y)\in{}F$, where $E$ and $F$ are relations, and
the roots of these rules are therefore either $(x,y)\in{}E$, or
$(x,y)\not\in{}E$. As a consequence, these rules cannot be applied to formulas
in which pairs are not explicit, such as $x\in{}E$ or $x\not\in{}E$ for
example. To deal with this problem, the idea is to add another proposition
rewrite rule for the product, i.e. the rule $R:x\in{}a\times{}b\rightarrow{}
\ex{}y\ex{}z\;(x=(y,z)\land{}(y,z)\in{}a\times{}b)$, which generates the two
following superdeduction rules:

\begin{center}
\begin{tabular}{cp{0.5cm}c}
\AXCm{x\in{}a\times{}b}\RLm{\times{}^*}
\UICm{x=(\epsilon_y,\epsilon_z),\epsilon_y\in{}a,\epsilon_z\in{}b}\DP &&
\AXCm{x\not\in{}a\times{b}}\RLm{\neg{\times}^*}
\UICm{x\neq(Y,Z)\brsep{}Y\not\in{}a\brsep{}Z\not\in{}b}\DP
\end{tabular}
\end{center}

where the first rule introduces
$\epsilon_y=\epsilon(y).(\ex{}z\;(x=(y,z)\land{}(y,z)\in{}a\times{}b))$, and
$\epsilon_z=\epsilon(z).
(x=(\epsilon_y,z)\land{}(\epsilon_y,z)\in{}a\times{}b))$.

For each relation $p$, we also need to add another proposition rewrite rule of
the form $R:x\in{}p\rightarrow{}\ex{}y\ex{}z\;(x=(y,z)\land{}(y,z)\in{}p)$, and
then generate the corresponding superdeduction rules. For example, for the
inverse relation, we obtain the rules as follows:

\begin{center}
\begin{tabular}{cp{0.5cm}c}
\AXCm{x\in{}a^{-1}}\RLm{{a^{-1}}^*}
\UICm{x=(\epsilon_y,\epsilon_z),(\epsilon_z,\epsilon_y){}\in{}a}\DP &&
\AXCm{x\not\in{}a^{-1}}\RLm{\neg{}{a^{-1}}^*}
\UICm{x\neq(Y,Z)\brsep{}(Z,Y)\not\in{}a}\DP
\end{tabular}
\end{center}

where the first rule introduces
$\epsilon_y=\epsilon(y).(\ex{}z\;(x=(y,z)\land{}(y,z)\in{}a^{-1}))$, and
$\epsilon_z=\epsilon(z).(x=(\epsilon_y,z)\land{}(\epsilon_y,z)\in{}a^{-1})$.

With these new rules, we can prove
$p\subseteq{}a\times{}b\implies{}x\in{}p\implies{}x\in{}(p^{-1})^{-1}$ in the
following way (we start the proof after the series of $\alpha_{\neg\implies}$
has been applied, and we do not detail the definitions of the involved
$\epsilon$-terms):

$$\rootAtTop
\AXCm{\begin{array}{c}
\vdots{}\\
\odot{}
\end{array}}
\UICm{X\not\in{}p}
\AXCm{\begin{array}{c}
\vdots{}\\
\odot{}
\end{array}}
\UICm{x\neq(Y,Z)}
\AXCm{\begin{array}{c}
\vdots{}\\
\odot{}
\end{array}}
\UICm{x\neq(Y,Z)}
\AXCm{\odot{}}\RLm{\odot_r}
\UICm{p\neq{}p}\RLm{\mathrm{pred}}
\BICm{(Y,Z)\not\in{}p}\RLm{\neg{}a^{-1}}
\UICm{(Z,Y)\not\in{}p^{-1}}\RLm{\neg{}{a^{-1}}^*}
\BICm{X=(\epsilon_y,\epsilon_z),\epsilon_y\in{}a,\epsilon_z\in{}b}
\RLm{\times{}^*}
\UICm{X\in{}a\times{}b}\RLm{\subseteq{}}
\BICm{p\subseteq{}a\times{}b,x\in{}p,x\not\in{}(p^{-1})^{-1}}\DP$$

where $X$, $Y$, and $Z$ are respectively instantiated by $x$, $\epsilon_y$, and
$\epsilon_z$.


\section{Implementation and Benchmarks}
\label{sec:bench}

The implementation of our extension of \zenon{} for the \bmth{} set theory
described in Sec.~\ref{sec:bset} has been possible thanks to the ability of
\zenon{} to extend its core of deductive rules to match specific requirements
like superdeduction. Concretely, this extension is an \ocaml{} file in which the
superdeduction rules of Figs.~\ref{fig:bset-sded1} and~\ref{fig:bset-sded2} are
implemented (about 2,000~lines of code). This file is loaded through
command-line options when \zenon{} is started, along with a \coq{} file
containing the translation of the superdeduction rules and used to generate
\coq{} proofs.

As said in the introduction, one of the main motivations for this extension of
\zenon{} for the \bmth{} set theory is to verify \bmth{} proof rules of
\atelierb{}~\cite{Atelier-B}, and in particular rules coming from the database
maintained by \siemens{}. Concretely, a rule is mainly a set formula with
guards, which are used to control the application of the rule (verifying that a
given hypothesis is in the context, for example). The verification of a rule has
been formally described in~\cite{JA13}, and consists in verifying that the rule
is well-typed, well-defined, and that it can be derived using the rules of the
\bmth{} proof system (see~\cite{B-Book}). Over the last few years, \siemens{}
has developed a formal and mechanized environment, named \bcare{}, which is
dedicated to the rule verification, and which relies in particular on a deep
embedding of the \bmth{} set theory in \coq{}, called \bcoq{}. In this
environment, our extension of \zenon{} is supposed to deal with the last step of
verification of a rule, i.e. the derivation of the rule within the \bmth{} proof
system.

\begin{figure}[t]
\includegraphics[width=0.9\textwidth]{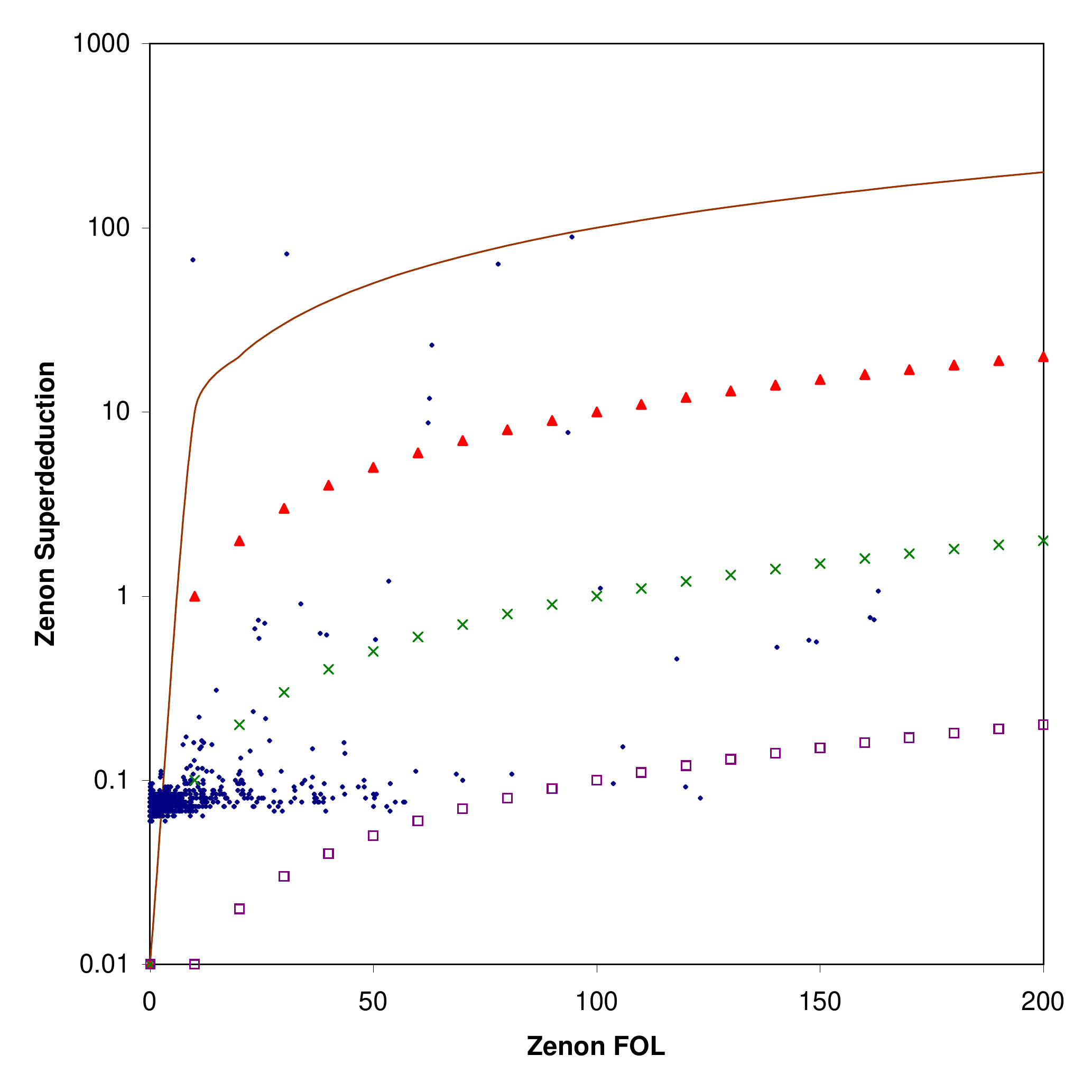}
\caption{Proof Time Comparative Benchmark}
\label{fig:bench1}
\end{figure}

\begin{figure}[t]
\includegraphics[width=0.9\textwidth]{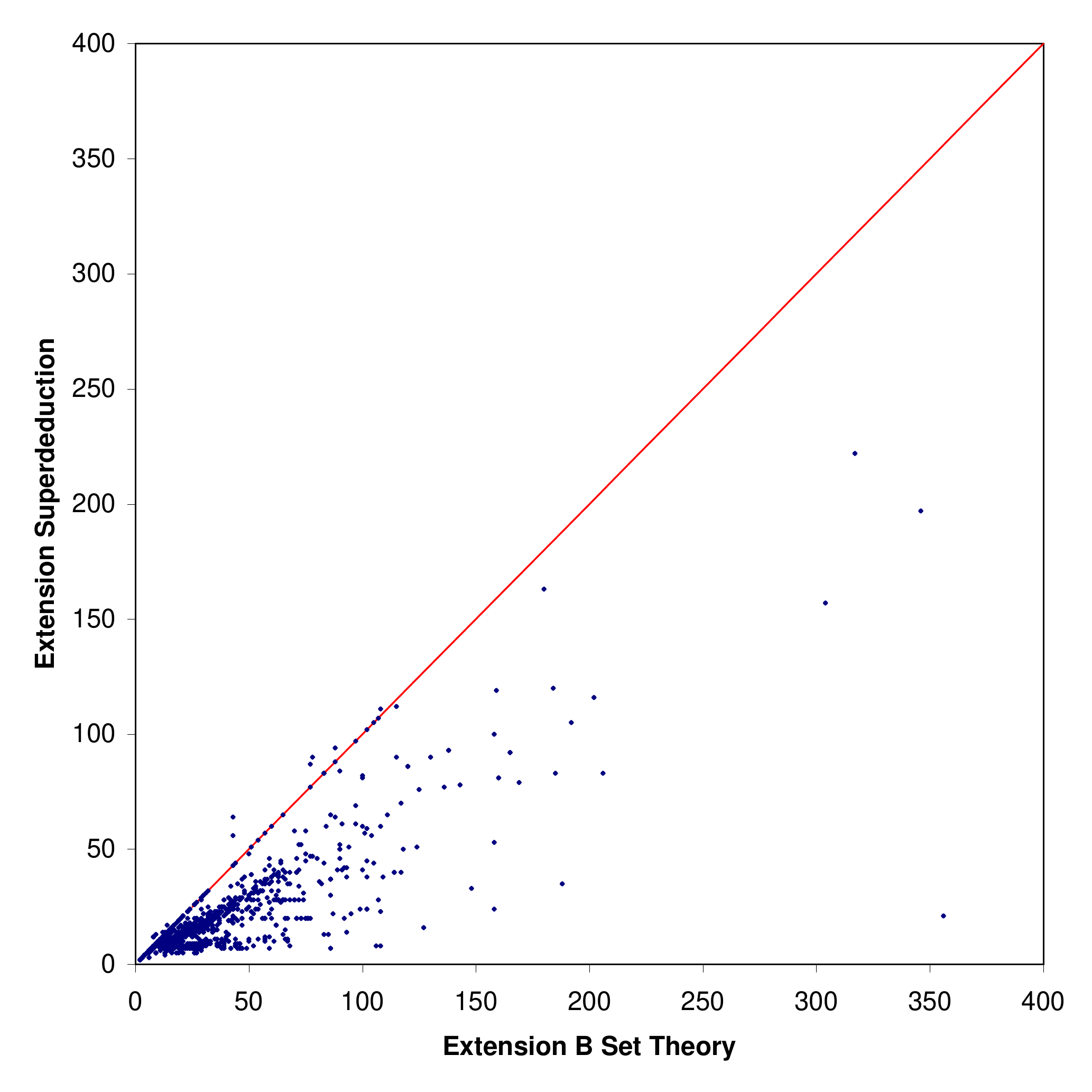}
\caption{Proof Size Comparative Benchmark}
\label{fig:bench2}
\end{figure}

Regarding benchmarks, we consider a selection of \bmth{} proof rules coming from
the database maintained by \siemens{}. This selection actually consists of
well-typed and well-defined rules, which involve all the \bmth{} set constructs
currently handled by the implementation of our extension of \zenon{}, i.e. all
the constructs of Chap.~2 of the \bbook{}~\cite{B-Book} until the override
construct (noted $\overridet$). This represents a subset of 1,397~rules, and we
propose two benchmarks whose results are gathered in Figs.~\ref{fig:bench1}
and~\ref{fig:bench2}.

The first benchmark aims to compare our extension of \zenon{} with the approach
coming from another experiment described in~\cite{JA13}. In this approach,
\zenon{} is simply used as a first order theorem prover (no set theory is
added), and the set formulas must be preliminarily normalized in order to obtain
first order logic formulas containing only the $\in$ set operator (which is
considered as an uninterpreted predicate symbol). This pre-normalization is
performed in \coq{} (within the \bcoq{} embedding), before calling
\zenon{}. Over the 1,397~selected rules, our extension of \zenon{} is able to
prove 1,340~rules (96\%), while our initial approach can deal with 1,145~rules
(82\%), which represents an increase of 195~rules (14\%). The difference of
proved rules mostly comes from ineffective parts of the implementation of our
initial approach, which still have to be optimized, and is not due to the
incompleteness of this approach (which is not pointed out over the considered
subset of rules).

The graph of Fig.~\ref{fig:bench1} presents a comparison of both approaches in
terms of proof time (run on an \intelc{}-2500K 3.30GHz/12GB computer) for a
subset of the 1,397~selected rules, where both approaches succeed in finding a
proof (the time measurement includes the compilation of \coq{} proofs generated
by \zenon{}), i.e. for 1,145~rules. In this figure, a point represents the
result for a rule, and the x/y-axes respectively correspond to the \zenon{}
approach with pre-normalization of the formulas and to our extension of \zenon{}
using superdeduction. For each point under the triangled curve, our extension
based on superdeduction is at least 10~times faster than our initial approach
with pre-normalization of the formulas; under the crossed curve, it is 100~times
faster, and under the squared curve, it is 1,000~times faster. We can observe
that there is a lot of rules for which the proof is~10 or 100~times faster with
superdeduction, and on average, the superdeduction proofs are obtained 67~times
faster (the best ratio is 1,540). These results are quite satisfactory in the
sense that our extension based on superdeduction solves the inefficiency issues
of our initial approach (mainly due to the pre-normalization, which is highly
time-consuming).

We propose a second benchmark whose purpose is to emphasize the proof size
speed-up offered by superdeduction. In this benchmark, the idea is to compare
our extension of \zenon{} using superdeduction with another extension of
\zenon{} for the \bmth{} set theory, where the proposition rewrite rules are not
computed into superdeduction rules, but just unfolded/folded (like in Prawitz's
initial approach~\cite{DP65}). The comparison consists in computing the number
of proof nodes of each proof generated by \zenon{}. We consider a subset of
1,340~rules, i.e. the rules for which both extensions succeed in finding a
proof. The results are summarized by the graph of Fig.~\ref{fig:bench2}, where a
point represents the result for a rule, and where the x/y-axes respectively
correspond to the extension without and with superdeduction. As can be seen, the
major part of proofs in superdeduction are shorter than the proofs where the
rules have not been computed and on average, the former have 1.6~times less
proof nodes than the latter (the best ratio is 6.25).

The previous benchmarks tend to show that the use of our extension of \zenon{}
for the \bmth{} set theory and based on superdeduction is quite effective in
practice, and very promising in terms of scalability. In particular, this could
be also applied to the verification of \bmth{} proof obligations (which involve
much larger formulas than \bmth{} proof rules).


\section{A Generic Implementation for First Order Theories}
\label{sec:first}

In this section, we present another extension of \zenon{} with superdeduction,
which is able to deal with any first order theory. In this extension, the theory
is analyzed to determine the axioms that can be turned into superdeduction
rules, and these superdeduction rules are automatically computed on the fly to
enrich the deductive kernel of \zenon{}. We also provide a comparative benchmark
coming from the TPTP library, which contains a large set of first order
problems, and which is usually used to test the implementations of automated
theorem provers.

\subsection{From Theories to Superdeduction Systems}

This extension of \zenon{} is actually a generalization of the previous one
dedicated to the \bmth{} set theory, where superdeduction rules are
automatically computed on the fly. In the previous extension, superdeduction
rules are hard-coded since the \bmth{} set theory is a higher order theory due
to one of the axioms of the theory (the comprehension scheme), and we have to
deal with this axiom specifically in the implementation of \zenon{}. Even though
some techniques exist to handle higher order theories as first order theories
(like the theory of classes, for example), a hard-coding of these theories may
be preferred as these techniques unfortunately tend to increase the entropy of
the proof search. In addition, in the previous extension, some of the
superdeduction rules must be manually generated as they must be shrewdly tuned
(ordering the several branches of the rules, for instance) to make the tool
efficient. The new extension of \zenon{} dealing with any first order theory has
been developed as a tool called \szen{}~\cite{Super-Zenon}, where each theory is
analyzed to determine the axioms that are candidates to be turned into
superdeduction rules. As said in Sec.~\ref{sec:sded}, axioms of the form
$\forall{}\bar{x}\;(P\Leftrightarrow{}\varphi{})$, where $P$ is atomic, can be
transformed, but we can actually deal with more axioms, in particular with
axioms of the form $\forall{}\bar{x}\;(P\Rightarrow{}\varphi{})$, which is
actually equivalent to perform polarized deduction modulo as introduced
in~\cite{GD02}. Here is the exhaustive list of axioms that can be handled by
\szen{}, as well as the corresponding superdeduction rules that can be generated
(in the following, $P$ and $P'$ are atomic, and $\varphi{}$ is an arbitrary
formula):

\begin{itemize}
\item Axiom of the form $\forall{}\bar{x}\;(P\Leftrightarrow{}\varphi{})$: we
consider the proposition rewrite rule $R:P\rightarrow{}\varphi{}$, and the two
superdeduction rules $R$ and $\neg{}R$ are generated;

\item Axiom of the form $\forall{}\bar{x}\;(P\Rightarrow{}P')$: we
consider the proposition rewrite rules $R:P\rightarrow{}P'$ and
$R':\neg{}P'\rightarrow{}\neg{}P$, and only the superdeduction rules $R$ and
$R'$ are generated;

\item Axiom of the form $\forall{}\bar{x}\;(P\Rightarrow{}\varphi{})$: we
consider the proposition rewrite rule $R:P\rightarrow{}\varphi{}$, and only the
superdeduction rule $R$ is generated;

\item Axiom of the form $\forall{}\bar{x}\;(\varphi{}\Rightarrow{}P)$: we
consider the proposition rewrite rule $R:\neg{}P\rightarrow{}\neg{}\varphi{}$,
and only the superdeduction rule $R$ is generated;

\item Axiom of the form $\forall{}\bar{x}\;P$: we consider the degenerated
proposition rewrite rule $R:\neg{}P\rightarrow{}\bot{}$, and only the
superdeduction rule $R$ is generated.
\end{itemize}

The axioms of the theory that are not of these forms are left as regular
axioms. An axiom that is of one of these forms is also left as a regular axiom
if the conclusion of one of the generated superdeduction rules (i.e. the top
formula of one of these rules) unifies with the conclusion of an already
computed superdeduction rule (in this case, the theory is actually
non-deterministic, and we try to minimize this source of non-determinism by
dividing these incriminated axioms among the sets of superdeduction rules and
regular axioms). An axiom that is of one of these forms is still left as a
regular axiom if $P$ is an equality (as we do not want to interfere with the
specific management of equality by the kernel of \zenon{}). Finally, for axioms
of the form $\forall{}\bar{x}\;(P\Rightarrow{}P')$, we also consider the
proposition rewrite rule that corresponds to the converse of the initial
formula; this actually allows us to keep cut-free completeness in this
particular case.

\subsection{Benchmark from the TPTP Library}

To assess the effectiveness of \szen{} compared to the regular version of
\zenon{}, we propose a benchmark that consists of problems coming from the TPTP
library~\cite{TPTP}. This library is a large collection of problems, which is
used to test the implementations of automated theorem provers in
particular. From this library, we consider a selection of problems that consists
of the set of non-clausal first order problems, i.e. the problems of the FOF
category in the TPTP library. The results of this experiment (run on an
\intelp{} 3.60GHz/32GB computer) are summarized in Tab.~\ref{tab:first}, where
both \zenon{} and \szen{} are called over the problems of the FOF category. As
can be observed, \szen{} is able to prove more problems than \zenon{} in the
whole FOF category with a significant rate (about 7\%). This rate becomes even
much better in some specific sub-categories of the FOF category, like in the SET
sub-category (about 37\%), which gathers problems of set theory. These results
in the SET sub-category tend to confirm the other results described in
Sec.~\ref{sec:bench} in the framework of the \bmth{} method, and also tend to
show that set theory is an appropriate theory to be handled by superdeduction
and by deduction modulo more generally. In addition to the problems of the FOF
category, \szen{} has been also applied to the counter-satisfiable problems of
the TPTP library, i.e. problems that are not valid, and \szen{} does not find
any proof for these problems, which allows us to have a relative confidence in
the correctness of the implementation of this tool.

\renewcommand{\arraystretch}{1.2}
\newcolumntype{C}{>{\centering}X}

\begin{table}[tb]
\begin{center}
\begin{tabularx}{\textwidth}{|C|C|C|}
\hline
\begin{tabular}{c}
TPTP\\
Category
\end{tabular} &
\zenon{} &
\szen{}\tabularnewline
\hline
\begin{tabular}{c}
FOF\\
6,644~problems
\end{tabular} &
1,646 & 
1,765 {\scriptsize (7.2\%)}\tabularnewline
\hline
\begin{tabular}{c}
SET\\
462~problems
\end{tabular} &
147 &
202 {\scriptsize (37.4\%)}\tabularnewline
\hline
\end{tabularx}
\end{center}
\caption{Experimental Results over the TPTP Library}
\label{tab:first}
\end{table}

\renewcommand{\arraystretch}{1}


\section{Superdeduction for Automated Deduction}
\label{sec:sauto}

Superdeduction can be seen as a generic method to integrate axiomatic first
order theories into deductive formal proof systems, like sequent calculus, as
well as into proof search methods that are very closed to deductive formal proof
systems, like the tableau method, which is a proof search method in sequent
calculus without cut. In particular, as seen previously, it has allowed us to
smoothly integrate the set theory of the \bmth{} method into the tableau method,
which has been implemented as an extension of the \zenon{} automated theorem
prover, and which has been used for the verification of \bmth{} proof rules. The
considered approach is actually so generic that it has been generalized into a
new extension of \zenon{}, called \szen{}, which is able to deal with any
axiomatic first order theory. Thus, any user that aims to perform automated
deduction in a given axiomatic first order theory must just provide the theory
to \szen{}, which is able to automatically integrate a part of this theory as
superdeduction rules. The approach is even more generic since it can be actually
implemented in any tool based on a tableau method. As for the application of
superdeduction to other proof search methods, such as resolution for example,
the approach is probably less straightforward since resolution proofs are
actually far from proofs in usual deductive formal proof systems, like sequent
calculus, which makes the introduction of superdeduction rules harder as these
rules are pure deduction rules. However, some work has been done in this
direction in the framework of deduction modulo~\cite{DA03} (whose superdeduction
may be considered as a variant), with a concrete development, called
\iproverm{}~\cite{GB11b}, implemented as an extension of \iprover{}~\cite{KK08},
which is a resolution-based automated theorem prover.


\section{Conclusion}

We have proposed a method that allows us to develop tableaux modulo theories
using superdeduction. This method has been presented in the framework of the
\zenon{} automated theorem prover, and applied to the set theory of the \bmth{}
method. This has allowed us to provide another prover to \atelierb{}, which can
be used to verify \bmth{} proof rules automatically. We have also proposed some
benchmarks using rules coming from the database maintained by \siemens{}. These
benchmarks have emphasized significant speed-ups both in terms of proof time and
proof size compared to previous and alternative approaches. Finally, we have
described another extension of \zenon{} with superdeduction, called \szen{},
which is able to deal with any first order theory. In this extension, the theory
is analyzed to determine the axioms that can be turned into superdeduction
rules, and these superdeduction rules are automatically computed on the fly to
enrich the deductive kernel of \zenon{}. A comparative benchmark that consists
of a large set of first order problems coming from the TPTP library has shown a
significant improvement of \szen{} over the regular version of \zenon{}.

As future work, we must extend our specific implementation realized for
verifying \bmth{} proof rules in order to deal with a larger set of rules coming
from the database maintained by \siemens{}. More precisely, the next step is to
consider functions, and the corresponding rules should point out the
incompleteness of our initial approach that consists in pre-normalizing the set
formulas. For instance, formulas such as
$\{~(x,y)~|~x=y~\}\in{}{a\fctpart{}a}\Rightarrow{}
\ex{}f\;(f\in{}{a\fctpart{}a}\land{}(b,b)\in{}f)$ should be proved by our
extension based on superdeduction, but not by our initial approach (since an
additional normalization is required once $f$ has been instantiated by the
comprehension set in hypothesis). We also plan to enhance the heuristic used by
\szen{} to transform axioms into superdeduction rules in order to increase the
improvement of \szen{} over \zenon{} in the framework of generic theories. In
addition, it should be worth building an heuristic with transformation rules
that preserves some important properties (from the automated deduction point of
view) of the initial axiomatic theories, such as cut-free completeness for
example.

\begin{ack}
Many thanks to G.~Burel and O.~Hermant for their detailed comments on this
paper, to G.~Dowek for seminal discussions about this work, and to D.~Doligez
for his help in the integration of superdeduction into \zenon{}.
\end{ack}

\bibliographystyle{abbrv}
\bibliography{biblio}

\end{document}